\documentclass{elsart1p}
\usepackage{graphicx}

\begin{document}
\begin{frontmatter}
\title{Low temperature ordering in easy-axis $S=1$ kagome and triangular lattice antiferromagnets}
\author{Kedar Damle}
\address{Department of Theoretical Physics,
Tata Institute of Fundamental Research,
1 Homi Bhabha Road, Mumbai 400005, India}
\date{March 15 2007}
\maketitle
\begin{abstract}
I review recent work in collaboration with T.~Senthil~\cite{Damle_Senthil}  on the low temperature phases of $S=1$
kagome lattice antiferromagnets in which there is sufficiently strongly single-ion anisotropy $D$
that dominates over the antiferromagnetic exchange $J$. Earlier results (in collaboration with D.~Heidarian~\cite{Heidarian_Damle}) that are relevant for the low temperature
physics of similar systems on the triangular lattice are also described briefly.
\end{abstract}

\end{frontmatter}
\section{Introduction}
In many electrically insulating magnetic materials, the basic interaction between the magnetic moments may be encoded
in terms of the antiferromagnetic exchange energy $E =  J \sum_{\langle i j \rangle} {\mathbf S}_i \cdot {\mathbf S}_j  \; ;J>0 $,
where $J$ is the exchange constant and the subscripts refer to pairs of nearest-neighbour moments. Here, the spins
${\mathbf S}$ are of course quantum-mechanical operators; however, for many purposes at not too low temperatures,
they can be usefully approximated by classical vectors of fixed length, particularly if the spin quantum number $S$ is $3/2$
or higher.

\begin{figure}
\begin{center}
 \includegraphics[width=3cm]{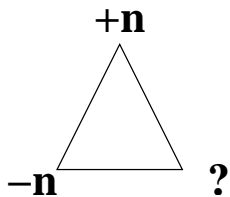}
 \caption{Three spins interacting antiferromagnetically with each other cannot satisfy the demands of all the exchange interactions}
\end{center}
 \label{triangles}
 \end{figure}

When the magnetic ions form a bipartite lattice, this energy is minimized by the so-called {\it Neel} state in which all spins
lie along a spontaneously chosen axis ${\bf n}$ and every spin points anti-parallel to its nearest neighbours [In two and higher
dimensions, this picture also gives an essentially correct caricature of the ground state of the full quantum
problem on a square or hypercubic lattice.]
When one talks of {\it frustrated} antiferromagnetism, one has in mind magnetic lattices with triangular motifs
in them. Clearly, the Neel (antiferromagnetic) state along any axis ${\bf n}$  is {\it frustrated} in the presence of such triangles, since there is no unique
way of satisfying all
the exchange interactions (Fig~\ref{triangles}). 

In many situations~\cite{exptreview1}, this results in a macroscopic degeneracy of {\it classical} minimum energy configurations.
At intermediate temperatures $T$ that are less than the exchange $J$, but are not small enough for the quantum mechanical nature of spins to matter, the spin correlations (measured,
say, by neutron scattering experiments) in the system simply reflect this macroscopic degeneracy, and can be modeled in a universal
way in terms of averages over an ensemble that gives equal weight to each of these minimum energy configurations~\cite{Moessnerreview}.
However, the ultimate fate of the system at very low temperatures is of course less universal, and depends sensitively on the
effects of quantum fluctuations and other (subdominant) interactions  acting in this subspace.

\begin{figure}
\begin{center}
\begin{tabular}{llr}
 \includegraphics[width=3.0cm]{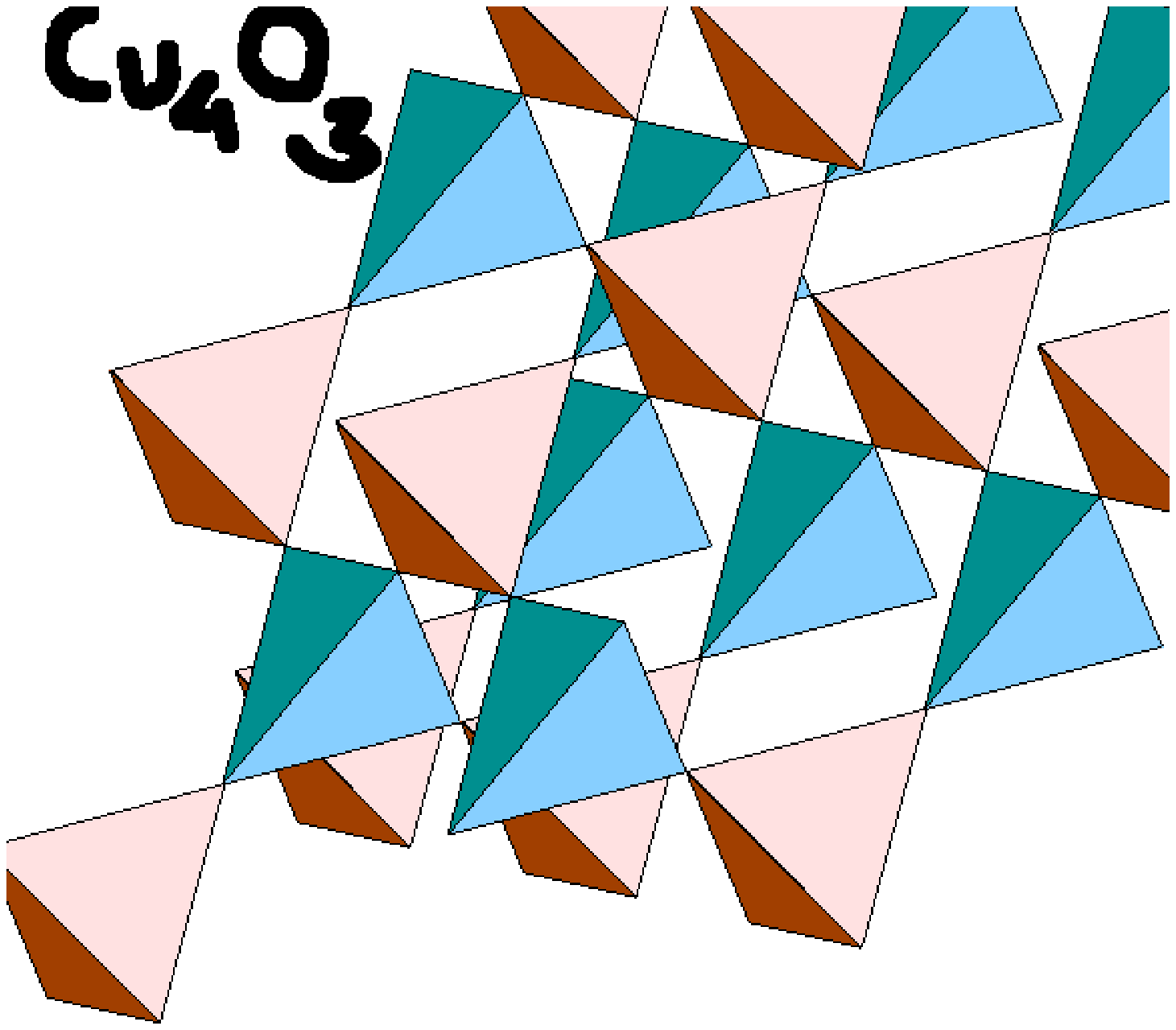} & \includegraphics[width=3.0cm]{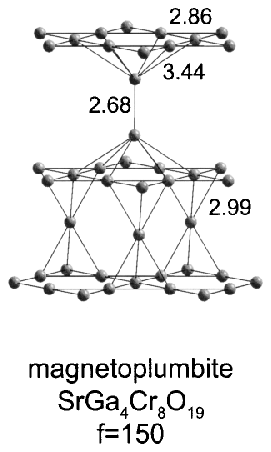} &  \includegraphics[width=3.0cm]{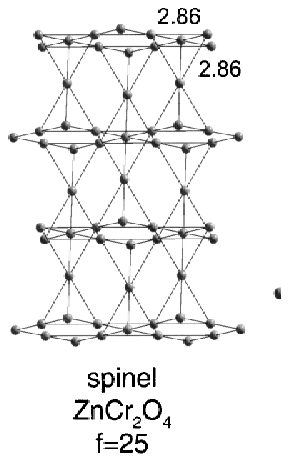}\\
 Paramelaconite&  Cr$^{3+}$ based & Cr$^{3+}$ based\\
\end{tabular} 
\caption{Some examples: On left is the pyrochlore structure made up of corner sharing tetrahedra whose centers form the diamond lattice,
while on the right is the layered kagome structure made up of corner sharing triangles whose centers form the honeycomb net (Figure
on the right is taken from Ref~{\protect \cite{otherkagome1}})}
\end{center}
\label{examples}
\end{figure}

Many examples of frustrated magnets are known (Fig~\ref{examples}). On the pyrochlore lattice, these
include the $Cu^{2+}$ based $S=1/2$ magnet paramelaconite~\cite{pyro1} and the $Cr^{3+}$ based $S=3/2$ magnets
CdCr$_2$O$_4$ and HgCr$_2$O$_4$~\cite{pyro2}. Several interesting examples have also been studied
on the kagome lattice---these include $Cu^{2+}$ based $S=1/2$
volborthite and other systems\cite{otherkagome0}, $Ni^{2+}$ based $S=1$ magnets Ni$_3$V$_2$O$_8$\cite{nickel1}, $Cr^{3+}$ based $S=3/2$ systems
\cite{otherkagome1}, and $Fe^{3+}$ based $S=5/2$ magnets Fe jarosite
\cite{otherkagome2}. 
	
While these nearly isotropic examples, particularly those with spin $S=1/2, 1$ pose interesting questions to the theorist regarding the true low temperature
state of an idealized quantum antiferromagnet on these frustrated lattices, there are many other equally interesting
examples in which anisotropy effects, particularly {\it single-ion anisotropy} (a term in the Hamiltonian of
the form $-D({\mathbf S} \cdot {\mathbf n})^2$, where ${\bf n}$ is the {\it easy-axis}) are strong, and sometimes dominant.

One such example in which anisotropy effects dominate is provided by the pyrochlore {\it spin ice}~\cite{Harris}  compound Ho$_2$Ti$_2$O$_7$  (Ho$^{3+}$, $(L+S)=8$),
in which the easy axes ${\mathbf n}$ point outward from center of each tetrahedron and $D \sim 50K$ is much
larger than {\it ferromagnetic} $J \sim 1K$---indeed, it is the geometry of the easy axes that {\it frustrates}
the ferromagnetic interaction in this case.

Another more recent example is the kagome compound Nd-langasite~\cite{Robert_etal,Bordet_etal}  Nd$_3$Ga$_5$SiO$_{14}$ (Nd$^{3+}$, $(L+S)=9/2$).
In this material, the dominant anisotropy term at low temperature is again of the easy-axis variety with the easy axes
now pointing perpendicular to the kagome planes.

Such a strong easy axis anisotropy allows one to use a  pseudo-spin-1/2 `Ising' description in which each spin has only
two allowed states ${\mathbf S} \cdot {\mathbf n} = \pm S$. The low temperature physics is then governed by the action of
quantum and thermal fluctuations within the low-energy manifold defined by the restriction of each spin to these allowed states.

\section{Easy-axis $S=1$ kagome magnet}
With this background, we consider the $S=1$ easy axis Kagome lattice antiferromagnet with Hamiltonian
\begin{equation}
H=H_0+H_1 \; ,
\end{equation}
where $H_0 = -D\sum_i(S_i^z)^2+J\sum_{ij}S_i^z S_j^z -bJ\sum_iS_i^z$,
$H_1 = J\sum_{ij} {\mathbf S}_i^{\perp} \cdot {\mathbf S}_j^{\perp}$, and the exchange $J$ is operative
only between nearest neighbour pairs of spins.

What is the physics for small $J/D$? This is best answered by asking for the perturbative effective Hamiltonian
that governs the small $J/D$ dynamics of pseudospin-$1/2$ (pauli-matrix) variables ${\mathbf \sigma}_i$ 
which represent the low-energy states of each spin via the correspondence $\sigma^z = \pm1 \leftrightarrow S^z=\pm1$
(note that in this language, the pseudo-spin $1/2$ raising operator $\sigma^{+}$ that flips a $\sigma^z=-1$ to a $\sigma^z=+1$ state
actually transforms the state with $S^z=-1$ to that with $S^z=+1$, completely bypassing the high-energy state $S^z=0$).
We expect that such a perturbative approach will be valid as long as collinear low temperature states are selected by
the anisotropy term.

\begin{figure}
\begin{center}
{\includegraphics[width=3.0cm]{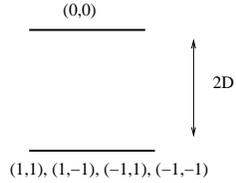}}
\caption{Level diagram of the two site system}
\end{center}
\label{twosite}
\end{figure}

To get a feel for how the perturbation theory in $J/D$ proceeds, it is enough to think about a simple two site
system (Fig~\ref{twosite}). Clearly, the $J S_1^z S_2^z$ term, being diagonal in $S_z$ basis results in an ${\mathcal O}(J)$ shift in the
energies of various $S^z$ eigenstates. On the other hand,  the $J{\mathbf S}_{1 \perp} \cdot {\mathbf S}_{2 \perp}$ term
produces (virtual) transitions to $(0,0)$ excited
state of the two spin system. These transitions are responsible for two distinct ${\mathcal O}(J^2/D)$ effects: one is a diagonal term
that represents an antiferromagnetic coupling $J_z$ between the pseudo-spins $\sigma^z$, while the other
is an off-diagonal pseudospin-1/2 exchange term $J_{\perp}$.

In the $S=1$ case, this two-site analysis generalizes easily to the full lattice, giving the leading order effective Hamiltonian
\begin{equation}
{\mathcal H}_{\mathrm eff} = \frac{J_z}{8}\sum_{\Delta} (\sigma^z_{\Delta} - b/2)^2 - \frac{J_\perp}{4}\sum_{ij}
(\sigma^+_i\sigma^-_j + \sigma^-_i\sigma^+_j)
\end{equation}
where $J_z \sim 4J +\dots$ is of course antiferromagnetic, while $J_\perp \sim J^2/D +\dots$ is
{\it ferromagnetic} in nature.
 With this in hand, our strategy below is to understand the phase diagram of ${\mathcal H}_{\mathrm eff}$ for general $J_z/J_{\perp}$
 and then specialize to the case $J_z \gg J_\perp$ to  draw conclusions regarding original $S=1$ problem.

One easy to understand regime is the regime in which the ferromagnetic in-plane interactions dominate, {\it i.e.} $J_\perp \gg J_z$.
The resulting ground state is of course a $x-y$ ferromagnet. A good (variational) wavefunction for
this ferromagnet is clearly $|\Psi \rangle_0 = \prod_i | \sigma^x_i = +1\rangle$ which polarizes all spins in the $x$ direction.
If we think of $\sigma^z = +1$ as presence of hard-core boson, and $\sigma^z = -1$ as vacancy, then this
is a {\it superfluid} state (in this particle language, $J_\perp$ is the particle hopping amplitude, while $J_z$ is nearest
neighbour repulsion). The in-plane ferromagnetic order in this state then corresponds to the {\it off-diagonal long-range order}
one expects in a superfluid, that is,
$\langle \sigma^{+}(r) \sigma^{-}(0)\rangle \rightarrow c_0^2 > 0$ as $r \rightarrow \infty$. 

\begin{figure}
\begin{center}
\includegraphics[width=3cm]{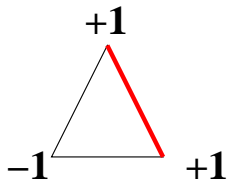}
\caption{A {\it minimally frustrated} triangle with exactly one frustrated bond in it}
\end{center}
\label{frustratedbond}
\end{figure}

With this in hand, we now ask what happens in this interaction dominated regime $J_z \gg J_\perp$?
Usually,  if interactions dominate physics of bosons on lattice, the particles localize in
some spatial arrangement that minimizes interaction energy, yielding a {\it bosonic Mott Insulator}.
However, in our problem, $J_z$ is frustrated, {\it i.e.} there is no unique lowest energy spatial
arrangement of bosons on the kagome lattice. Instead there is a vast degeneracy of minimally frustrated configurations:
All configurations
with exactly one {\it frustrated} bond in each triangle serve equally well to minimize the classical interaction
energy $J_z$ (Fig~\ref{frustratedbond}).

To make progress in such a situation, we appeal to ideas developed earlier in work on triangular lattice supersolids~\cite{Dhar,Melko_etal,Heidarian_Damle}. These ideas rely on the observation 
that the ground state lives entirely in minimum frustration subspace in the  $J_z/J_\perp \rightarrow \infty$~\cite{foot}.
However, the kinetic energy term would still prefer a superfluid state. A good way to reconcile these features is to simply
project the superfluid wavefunction $|\Psi\rangle_0$ to the minimally frustrated subspace, {\it i.e.} consider
the variational wavefunction $|\Psi\rangle_\infty = {\mathcal P}_g |\Psi \rangle_{0}$, where ${\mathcal P}_g$ is
the appropriate projection operator. 

In more pedestrian terms, this simply amounts to writing out $|\Psi \rangle_0$ in the $\sigma^z$ basis and keeping
only those terms in which the $\sigma^z$ configuration contains no triangle with more than one frustrated bond. Since
all these configurations enter the wavefunction with amplitude one,
$|\Psi\rangle_\infty$ is thus the equal amplitude superposition of all possible minimally frustrated configurations, {\it i.e.} all $T=0$ states of {\it classical} Ising model on Kagome lattice.

What is the rationale behind this wavefunction? The answer is simple: First of all,
the wavefunction minimizes the potential energy by construction. At the same time, the minimum frustration
subspace admits considerable density fluctuations (note that a minimally frustrated triangle can have either two sites occupied
and one unoccupied or two sites unoccupied and one site occupied). These density fluctuations suggest that
the wavefunction preserves the superfluid character of the unprojected state, and indeed, one can check
that $\langle \sigma^{+}(r) \sigma^{-}(0)\rangle_\infty \rightarrow c_\infty^2$ as $r \rightarrow \infty$, with $c_\infty^2 < c_0^2$
but non-zero. This superfluid character implies that the kinetic energy gain is also considerably substantial in this
wavefunction, allowing us to conclude that it does indeed provide a good variational description of the large $J_z$
physics.

One immediate consequence of this reasoning is that the the ferromagnetism (or off-diagonal long-range order in boson language) survives all the way to $J_z/J_\perp \rightarrow \infty$. What about $\sigma^z$ correlations?
From the explicit construction above, it is clear that our variational wavefunction {\it inherits} equal time $\sigma^z$ correlations of the classical Kagome lattice Ising model
at $T=0$. Borrowing from the work of Huse and Rutenberg~\cite{Huse_Rutenberg} on this classical model, we
then have $\langle \sigma^z(r) \sigma^z(0) \rangle_{\infty} \sim e^{-r/\xi}$,  that is, the $\sigma^z$ correlator is short-ranged
and there is no diagonal long-range order.

What are the implications of this variational line of thought when it comes to the anisotropic $S=1$ magnet we started
out with? Clearly, $\langle S^z(r) S^z(0) \rangle_{\infty} \sim e^{-r/\xi}$.  In addition, the transverse components ${\bf S}_\perp$
of the spin also have purely short ranged correlators, as may be checked easily within our variational framework. There is thus
absolutely no spin ordering in the large $D/J$ limit. However, the square of the transverse spin components {\it do}
order as is clear from the operator correspondence $(S^{+})^2 \sim \sigma^+$.
In other words, we expect $\langle (S^{+})^2(r) (S^{-})^2(0)\rangle_\infty \rightarrow c_{\infty}^2$ as $r \rightarrow \infty$.
Following the terminology of early proposals for the ordering of the square of the spin~\cite{Andreev_Grishchuk,Gorkov_Sokol,Chandra_Coleman}, we
dub this `spin-nematic' order.

 This is of course a $T=0$ result. For $T>0$, quasi-long range order in $(S^{+})^2$ will survive, in that there will
 be a non-zero stiffness to twists in the phase of the nematic order parameter and a corresponding propagating `nematic sound' mode
 (which is of course in direct correspondence with the propagating sound mode of the quasi-long range ordered Kosterlitz-Thouless superfluid 
 phase).

In terms of experimental signatures, we expect a finite linear magnetic susceptibility for magnetic fields both
parallel to, and perpendicular to, the easy axis. Also, the spin structure factor, as measured neutron scattering
experiments, will have no signs of any long-range order. However, there will be a  $T^2$ low temperature specific heat,
which is a direct consequence of the linearly dispersing mode of the nematic.

How does a finite field $B_z = Jb$ along the easy axis affect the physics?
Clearly,  the spin nematic is stable for small $b$, and the magnetization along the $z$ direction increases
smoothly in response to the field while preserving the nematic order.  However, as the field is increased further, one
expects that there is a transition to a magnetization plateau state with magnetization $m=1/3$ in units of the
saturation magnetization.

This may be understood quite easily by starting with the $J_\perp =0$ classical limit
of the pseudo-spin effective Hamiltonian. In this extreme limit, any non-zero $b$ immediately enforces a strong
$2$:$1$ constraint on the minimum energy configurations: Two spins in every triangle point must point up along
the field, while one must point down, {\it i.e.} antiparallel to the field.\begin{figure}
\begin{center}
\includegraphics[width=8cm]{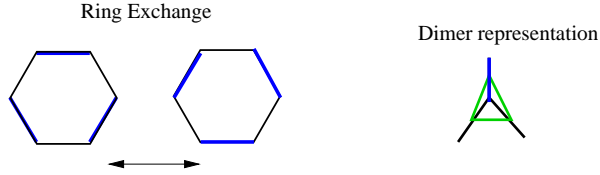}
\caption{ The small in-plane ferromagnetic exchange coupling induces a {\it ring exchange}
kinetic energy term in the dimer representation of the low-energy manifold of states at one-third magnetization (the blue dimer passes through the only
spin that is anti-parallel to the applied field)}
\end{center}
\label{ringexchange}
\end{figure}

Naturally, a small non-zero $J_\perp$ treated within perturbation theory then induces quantum fluctuations within this low-energy manifold, and the detailed properties of the resulting magnetization plateau state are thus controlled by the nature of the resulting effective Hamiltonian
that acts within this subspace.
This dynamics is best characterized by noting that each configuration in this low-energy subspace corresponds uniquely
to a dimer cover of the underlying honeycomb lattice whose links pass through the kagome lattice sites, with every down spin
associated with the presence of a dimer on the corresponding link of this honeycomb lattice.

\begin{figure}
\begin{center}
\includegraphics[width=6cm]{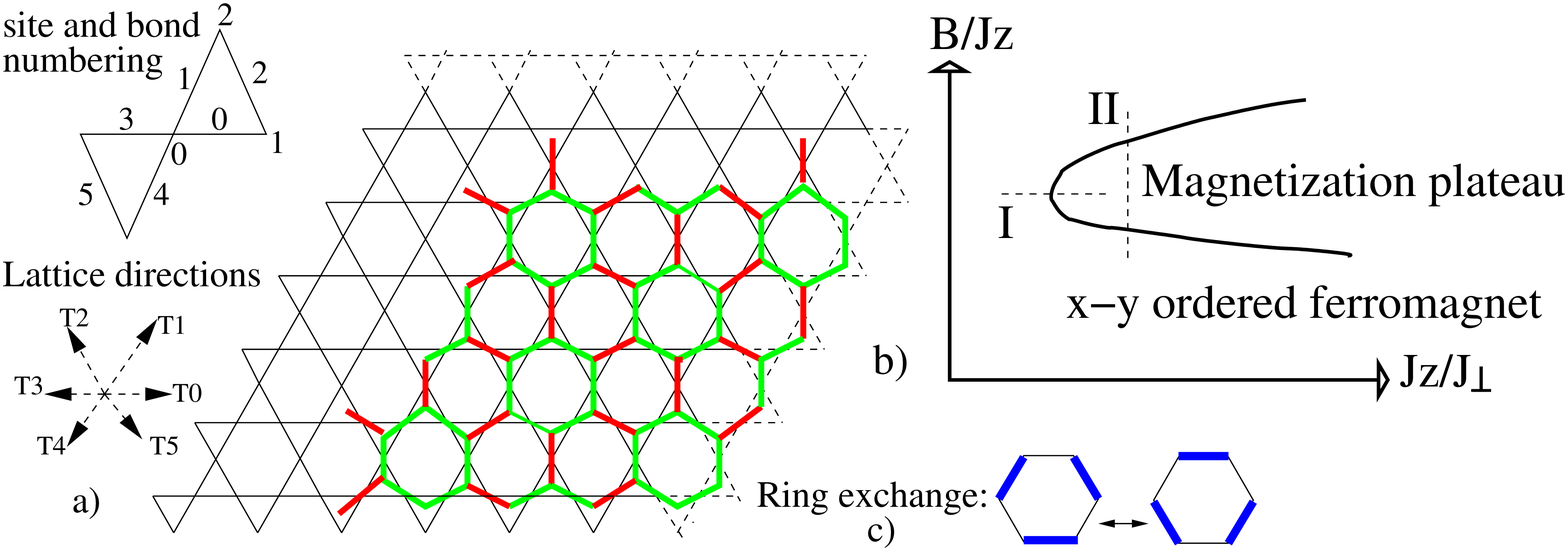}
\caption{ (From Ref~{\protect \cite{Damle_Senthil}})(a) Periodic Kagome lattice and the honeycomb net whose
bonds pass through the kagome sites. In the plaquette ordered state,
red honeycomb edges have no dimer ($\sigma^z=+1$), while green hexagons resonate via
the ring-exchange process (shown in (c)). In the alternate columnar state at the same wavevector, dimers
cover all red edges ($\sigma^z=-1$) but not green ones. Indirect evidence suggests that the plaquette state
is realized on the $m=1/3$ plateau. (b) Schematic phase diagram, showing two possibly
different phase transitions associated with the onset of the plateau---of these, the magnetic field
driven transition is of course readily accessible in experiments.}
\end{center}
\label{fig1kag}
\end{figure}

To leading order in $J_\perp/J_z$, the
effective Hamiltonian then consists of a ring-exchange kinetic energy term that allows every {\it flippable}
plaquette to resonate between its two allowed configurations (Fig~\ref{ringexchange}). Quantum dimer models of this
type tend to have lattice-symmetry broken `dimer-crystal' ground states~\cite{Sachdev_Vojta}, so that is what we expect on the magnetization
plateau. Detailed numerical work indeed confirms this expectation, and provides a picture of the resulting
state (see Fig~\ref{fig1kag}). In the original anisotropic $S=1$ problem, this of course corresponds
to a spin-density wave state in which the $S^z$ are `frozen' in a lattice-symmetry breaking pattern with distintive
bragg peaks that may be seen in neutron scattering experiments.

This completes our story on the kagome lattice. Two final comments are in order before we close: Firstly, arguments
and analyses entirely analogous to those described above, taken in conjunction with the original results of Ref~\cite{Heidarian_Damle,Melko_etal,Wessel_Troyer} and subsequent numerical studies of Ref~\cite{Prokofiev}, lead us to the conclusion that the large $D/J$
state of the anisotropy dominated $S=1$ triangular lattice antiferromagnet is a spin-nematic state {\it with
co-existing spin-density wave order}.
And finally, the transition between the spin-nematic and the spin-density wave state on the kagome lattice is also of considerable interest~\cite{Sengupta_etal,Isakov_etal}, but falls well outside the limited purview of the present review.

\section{Acknowledgements}
I would like
to acknowledge my collaborators D.~Heidarian and T.~Senthil, as well as L.~Balents, D.~Dhar, A.~Paramekanti,and A.~Vishwanath, for insightful comments
and useful discussions. Support from a Ramanujan Fellowship (DST), and
computational resources of TIFR are also gratefully acknowledged.


\begin{thebibliography}{999}
\bibitem{foot} For finite
but large $J_z/J_\perp$, a more precise statement is that the ground state is unitarily equivalent to some linear combination
of states in the minimally frustrated subspace. In other words, the physical ground state can be related to
a state in the minimally frustrated subspace by a unitary transformation that can be calculated order by
order in perturbation theory in $J_\perp/J_z$; this distinction does not matter at the level of our discussion here, but
can play a role in more detailed calculations.
\end{thebibliography}
\end{document}